\documentclass[conference]{IEEEtran}
\AtBeginDocument{%
  }

\usepackage[english]{babel}
\usepackage{todonotes}
\usepackage{url}
\usepackage{xcolor}
\usepackage{xspace}
\usepackage{float}
\usepackage{subcaption}
 
\usepackage{amsmath}
\usepackage{graphicx}

\newcommand{\eg}{\emph{e.g.,}\xspace}

\newcommand{\etal}{\emph{et~al.}\xspace}

\begin{document}

\title{Toward an Architectural Blueprint to Observe Sustainability in and by Software Systems}

\author{
Klervie Toczé\textsuperscript{1},
Andrei Dragomir\textsuperscript{1},
Vincenzo Stoico,
and Patricia Lago\\
Vrije Universiteit Amsterdam, Amsterdam, The Netherlands\\
Email: k.m.tocze@vu.nl, a.dragomir@student.vu.nl, \{v.stoico, p.lago\}@vu.nl \\
\textsuperscript{1}Klervie Toczé and Andrei Dragomir contributed equally.}

\maketitle

\begin{abstract}
Enabling observability in software systems brings many benefits. It can, for example, ease the identification of issues or the implementation of improvements.
It is especially critical to be able to observe sustainability-related dimensions of systems to know and mitigate their impact. 
However, adding observability to a system, especially related to software sustainability, requires technical knowledge that may not be available in every project that would benefit from it. 

In this work, we propose an architectural blueprint along with its deployment code that can be used to facilitate the addition of observability in software systems. As a special case, it includes measuring the energy consumption of software. This toolkit provides support in defining which components are necessary for a given use case and for structuring their deployment. Moreover, we exemplify the addition of observability in two different use cases. 
\end{abstract}

\begin{IEEEkeywords}
Observability, Monitoring, Dashboard
\end{IEEEkeywords}

\section{Introduction}
Software systems benefit from being monitored to decide whether their performance is satisfactory or not, and how they can be improved.
Therefore, an important property of software systems is \textit{observability}. 
Observability states how much stakeholders of a software system can understand the internals of the application using exposed outputs \cite{newman2021building}.
Additionally, software systems can be used to \textit{enable} observability in other types of systems, such as 
farms, 
manufacturing plants, and transportation networks, by collecting, processing, and analyzing data to monitor their behavior and improve performance, reliability, and sustainability \cite{almalki2021low}.

Although sustainability has become a central topic in software engineering discussions, there remains a notable gap: few approaches bridge sustainability metrics and observability, especially when it comes to distributed software systems, as well as in monitoring systems beyond software, including physical and industrial domains.
Existing performance observability literature focuses almost exclusively on traditional indicators such as latency, throughput, and resource utilization.
There is a need to bridge this gap while also collecting sustainability-related measurements that provide software stakeholders with the necessary data to support it.
To expose such metrics, software engineers need to implement a system to monitor, collect, and effectively communicate these metrics \cite{almalki2021low}.

However, enabling observability means deploying a comprehensive set of tools, ranging from tools to collect the required metrics to tools for visualizing them \cite{mayer2017dashboard}. There are two challenges here: 1) knowing which tool is relevant for each part and how each of them work, and 2) knowing how to connect all these tools. This knowledge may not exist in every project team using software for monitoring and acquiring it can be costly, especially if the core of the project is not about software. 
It is thus needed to provide guidance and tools to ease the deployment of observability stacks, especially addressing the second challenge so that the project team can focus their efforts on the project-specific deployment needs. 
If this deployment is made easier, a wider range of systems could benefit from observability. 

We propose an architectural blueprint to guide software practitioners (SPs) in the design and implementation of a software stack enabling domain-agnostic observability of sustainability-metrics.
The resulting observability stack is meant for SPs with intermediate technical knowledge 
who lack the experience of deploying a full stack, but are able to select the individual components and manage them once they are in place.

\textit{Why not just directly use AI for this?} While generative AI is increasingly applied to deployment tasks, its effectiveness depends on well-documented components. Subtle configuration differences across services are difficult for AI to capture when generating manifests from scratch. Once the general structure is in place, which is what this works is focusing on, AI may be used to do the tailoring of components. 

To summarize, we contribute: (i) An architectural blueprint enabling observability in software systems together with the corresponding code repository  to realize an observability stack, including energy-consumption metrics \cite{software-monitoring}, (ii) Two real sustainability-focused use cases illustrating the result when using the blueprint in practice,  (iii) A discussion of existing challenges and potential solutions.

\section{Related Work}
\label{sec:relatedworks}

Observability tools, and dashboards in particular, are commonly reported to be used in scientific literature for various use cases. However, current literature stop at merely stating that a dashboard tool is used, without providing further details on its implementation (\eg ~\cite{WOLFERT2022,FREITAS2025,Scipioni2009,Vannieuwenhuyze2020}).
This hinders reproducibility and re-use.

Beyond individual dashboards, a wide range of openly available observability tools exist for monitoring resource usage and energy consumption in software systems (\eg ~\cite{GMT_tool,GreenFrame_tool,CCF_tool}). While many of these tools are mature and well engineered, they are often presented as comprehensive solutions that are oriented towards Software Life Cycle Assessment (SLCA) or as tightly integrated platforms. As a result, SPs, especially those with limited prior expertise, are frequently left without clear methodological guidance on how to design and implement a monitoring environment tailored to their own application rather than adopting a fixed end to end solution. To address this gap, our blueprint and accompanying examples provide a structured methodology for assembling a modular observability stack using readily available, well documented, and highly configurable tools. As an illustrative example, CEEMS~\cite{CEEMS_tool} offers a resource manager agnostic energy and emissions monitoring stack that can be naturally integrated within the type of monitoring environment supported by our blueprint.

To lower the technical barrier to create observability tools (OTs) such as dashboards, tools have been created to ease for the novice OT creator. For example, Elias and Bezerianos~\cite{Elias_ExplorationViews} propose and evaluate a graphical tool for Business Intelligence dashboards, including recommendations for data mapping and visual templates. Bellini \etal~\cite{BELLINI_Snap4City} also use a graphical way to create dashboards for smart cities, to provide a solution with no or low coding as the OT creator may not have coding skills. Another approach is used by Clements~\cite{Clements_ExcelToShiny}, who created a package for automatically creating R dashboards from Excel spreadsheets. 
These works emphasize the need to provide tools adapted to the technical knowledge of the OT creator, to make dashboard creation accessible to a wider range of OT creators. Our work differs in two ways: we want to ease creating the full observability stack (and not the dashboard only), and we target intermediate OT creators, namely those with some technical experience. 

\section{Architectural Blueprint}
\label{sec:blueprint}

\begin{figure}[ht]
    \centering
    \includegraphics[width=\linewidth]{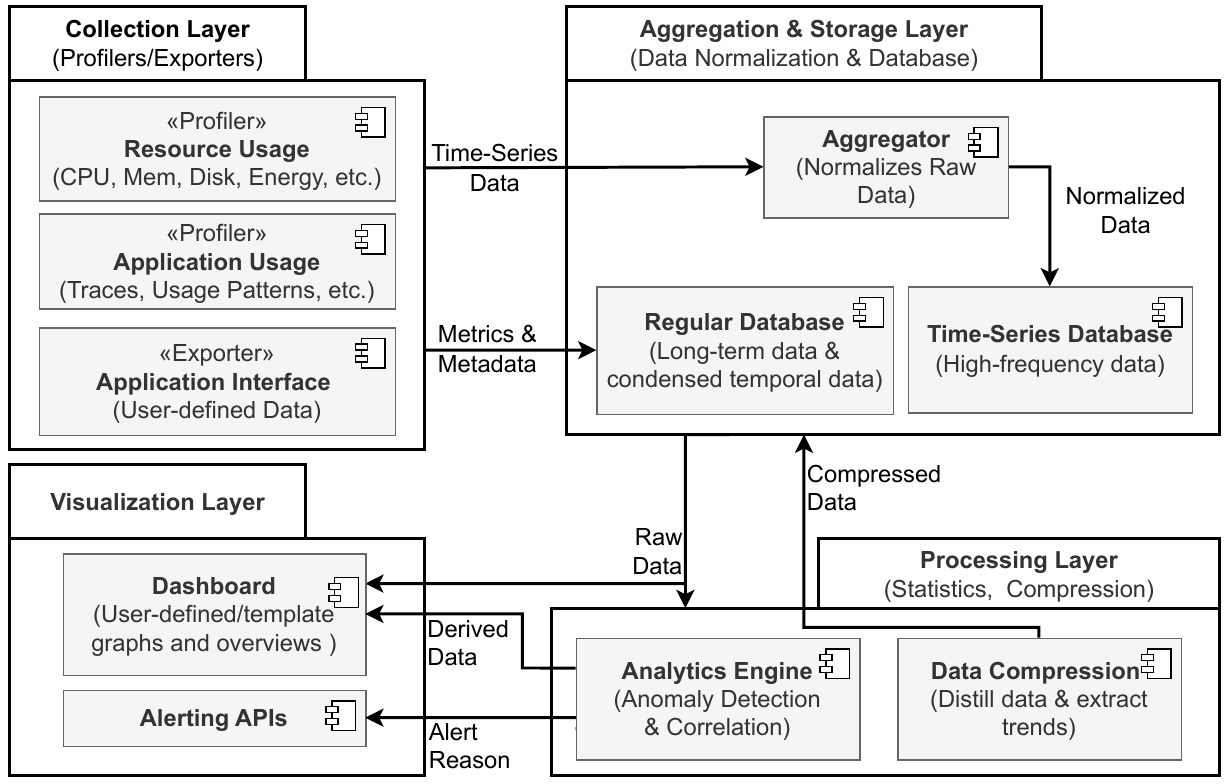}
    \caption{Architectural Blueprint}
    \label{fig:blueprint}
\end{figure}

Figure \ref{fig:blueprint} describes the main components and their relationships within our 
architecture.
This figure serves as a \textit{blueprint} that SPs can reuse and adapt to different systems and use cases. 
The blueprint emerged from our experience rather than a systematic design process, and the depicted elements are grounded in both literature and our use cases (see Section \ref{sec:usecases}).
The blueprint is organized in layers, each fulfilling a distinct role: collection, aggregation and storage, processing, and visualization.
The components follow a pipe-and-filters pattern, with data managed and passed sequentially until reaching the user-facing interface.
Figure \ref{fig:blueprint} should be read starting from the top left.

The data flow begins with the \textbf{Collection} layer, which profiles metrics of interest in a (software) system.
This includes resource and energy usage as well as application-specific data such as traces, usage monitoring, and patterns directly generated by end users.
Metrics can be captured at various granularities, from machine to function level, while user-generated data may range from interaction logs to domain-specific information (\eg purchase statistics or feature usage).
Collected data is transmitted in raw form, typically timestamped values, to the next layer via push or pull interfaces.

The \textbf{Aggregation \& Storage} layer consolidates this data. The \textit{Aggregator} gathers and \textit{normalizes} heterogeneous inputs into a common representation.
Normalization ensures comparability across profilers, \eg aligning CPU time units or structuring user interaction logs.
Push-based storage systems may embed aggregation, while other cases require 
an independent module to collect and organize user- or system-level data.
Data is stored in two databases: a \textit{time-series database} for high-frequency metrics such as resource usage, and a \textit{regular} \textit{database} for non time-series data, condensed statistics, and user or system metadata.
This distinction enables both efficient temporal queries and contextual long-term storage.

The \textbf{Processing} layer retrieves data through the \textit{Analytics Engine}, which supports all kinds of data analyzes. 
This includes, \eg anomaly detection, correlation, and root-cause reasoning by combining time-series with metadata.
It also performs compression, distilling raw time-series into compact summaries to optimize storage.
This establishes a continuous loop with the aggregation \& storage layer: gather, store, distill, and clear.
 Here, both system- and user-level data are processed, preparing tailored outputs for different stakeholders.

The \textbf{Visualization} layer provides the user-facing interface.
The \textit{Dashboard} integrates tracing, correlation, and monitoring views, supporting visualization of, \eg collected data 
and aggregated metrics.
Alerts are surfaced at this layer as they constitute the primary point of end-user interaction, delivered through webhooks or equivalent mechanisms.
Different audiences may receive distinct perspectives: administrators may access holistic system overviews, 
whereas individual users see only their own data and analytics.

\subsection{From Blueprint to Implementation}

We support the transition from blueprint to implementation through a code repository with templates for modular components and deployment instructions \cite{software-monitoring}.
This toolkit targets project members familiar with the components but needing guidance to deploy them as an integrated observability stack.

SPs start by identifying the blueprint components relevant to their specific use case across the different layers (\eg collection, storage, visualization). 
After selecting these components, they can customize the corresponding configuration files in the repository to match the desired setup.
Environment variables, including database and admin credentials, must be defined in a \texttt{.env} file, for which an example template is provided in the repository.
Deployment can then be carried out either by launching individual services using their dedicated deployment files or by merging multiple service definitions into a single \texttt{deployment.yml}.
The latter approach is recommended, as it simplifies customization and ensures all required services are started together.
After deployment, SPs may further tailor configurations, such as scrape targets or data sources, depending on the services involved. 
This process supports both system-wide observability views (for administrators) and user-specific analytics.  
For detailed instructions, examples, and deployment templates, please refer to~\cite{software-monitoring}.

\subsection{Challenges}
\label{sec:challenges}

The implementation of the proposed architectural blueprint requires consideration of several key aspects.

\textit{Not all the layers described in the blueprint are needed}.
While it is designed to capture diverse software characteristics, only the collection and the visualization layers are mandatory, as they provide the minimum needed to profile an application and offer insights.
For example, SPs may focus only on energy usage variation under workload. Implementing an observability stack can pose challenges related to analysis granularity, data sensitivity, storage, transmission, and visualization.

\textit{Observer effect:} A complex observability stack may require running several profiles in parallel to collect different types of data about the software system under observation. 
This process, which takes place in the Collection layer, consumes additional computing resources, potentially introducing overhead that can influence and bias the metrics of interest.
N{\~o}u \etal \cite{nou2025investigating} investigate how to trace impacts performance in microservices and serverless setups.
They show that tracing with OpenTelemetry and Elastic APM can slow down the throughput by up to 80\%, and latency can increase by 175\%.
The authors point out further challenges with the setup/initialization of the tracing system, data mole generated by fine-grained profiling, and data transmission.
Possible mitigation strategies involve balancing the trade-off between overhead and data granularity, batching data collection, and offloading processing to external machines.
The latter case can introduce burden on the network, especially if real-time data is needed.
\cite{traini2024vamp} tackle the problem by providing a system to visualize multiple end-to-end microservice requests, thus reducing data size when reaching the Visualization layer.

\textit{Security:} It is a concern affecting all layers of the observability stack.
Collected data may contain sensitive information about users or the system that should not be exposed.
Aghili \etal \cite{aghili2024empirical} discuss several cases where security threats arise in software logs, including the accidental disclosure of personal information and quasi-identifiers. 
This threat can be partially mitigated through anonymization or sanitization.

\textit{Aggregation vs. Granularity in Monitoring:} 
In software performance and energy analysis, aggregated data (such as average latency or average energy per request) are commonly stored and displayed.
Aggregation helps reduce storage requirements and simplifies real-time monitoring.
This approach can obscure important variations and hide information needed to diagnose unexpected issues.
The Visualization layer may report only the average latency for a software system undergoing 1000 request/minute, for example 200 ms, a value that satisfies extra-functional requirements.
However, examining disaggregated data (e.g., per-request), may reveal a different latency per request showing a bottleneck during processing.
Within that same minute, 950 requests could have completed in 100 ms, while 50 requests suddenly spiked to 3 seconds due to a temporary failure.
Yu \etal \cite{yu2023tracerank} in the context of root-cause analysis, which is usually done in Processing layer. 
Fault propagation due to a root cause service can cause surge in latency, which can influence the performance of multiple services.
Root cause analysis can be facilitated by automatically processing disaggregated data, since aggregation may hide the actual source of the fault.
However, storing every kind of data can overwhelm the database leading to huge memory consumption, slow queries, and high costs.

\textit{Dashboard Design:} The Visualization layer plays a vital role in enabling OT users to extract meaningful insights from collected data and to detect anomalies within software systems.
However, visualization also brings challenges related to interpretation and configurability.
Indeed, without proper guidance, OT users may struggle to interpret dashboards effectively.
Therefore, clear documentation and comprehensive guidelines are essential to help OT users create and understand visual dashboards.
Allowing OT users to build custom visualizations tailored to their specific metrics of interest can also be highly beneficial, particularly for non-IT OT users.
This process can be supported through the provision of predefined dashboard templates for specific use cases, e.g., a template focusing solely on sustainability metrics.
This feature is well-known and supported by tools for interactive and customizable dashboard creation like Grafana.

\section{Use Cases}
\label{sec:usecases}

We illustrate the use of our toolkit in two different scenarios: one focused on monitoring urban agriculture initiatives as part of a EU project, and another within our research group’s High-Performance Computing (HPC) lab, where it is used to track energy consumption and resource utilization of various software workloads. These workloads range from distributed applications to single processes, and the collected data serve as a basis for research and analysis. 

\subsection{Urban Agriculture Monitoring}

Our first use case is the Feed4Food project~\cite{Feed4Foodwebsite}. 
In this project, an observability tool including data gathering and dashboard functionalities is supporting three different urban agriculture setups called Living Labs (LL) that are developing urban gardens with a common goal (sustainable, inclusive and healthy urban food environments) but in different contexts (\eg size, location, and type of LL gardeners). 

Most of the data collected in this project comes from humans (\eg harvest, irrigation, event attendance) and some can be collected by software measurement tools (\eg CPU and energy use from the machine running the server), hence requiring the use of two different databases. 
Regarding data processing and visualization, there is a need to provide several visualization alternatives: 1) the LL gardeners want to see direct information about their own piece of land (\eg using the dashboard as a light bookkeeping tool), 2) the LL managers want to see how the LL performs with regards to defined targets, 3) the project partners want to learn and extract guidelines for future LLs based on the performance of all three LLs. Therefore, there is a need for three different types of visualizations: 1) a straightforward view on the data collected, 2) a view of key performance indicators derived from  the 
data collected, and 3) a highly flexible visualization option enabling researchers to ``play around'' with the data as needed. In addition, different analyzes need to be implemented to be able to get the data that should be visualized in 2) and 3).  

\begin{figure}
    \centering
   
    \includegraphics[width=0.85\linewidth]{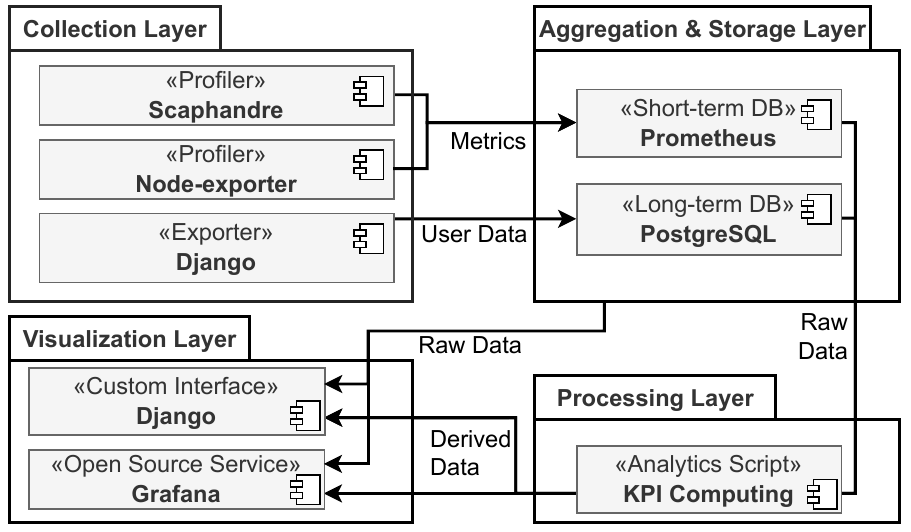}
    \caption{The Feed4Food stack}
    \label{fig:feed4foodStack}
\end{figure}

The specifics of the current architecture are illustrated in Figure \ref{fig:feed4foodStack}. The specificity of this use case is that part of the system observed (the gardens) is not a software system and that data for this part has to be manually gathered.  
The use case code is available 
on GitHub~\cite{Feed4Foodgithub}.  

\subsection{Distributed Energy and Resource Profiling}
Our second use case is situated in the GreenLab, our 
HPC laboratory \cite{malavolta2024ten}. 
There, the observability tool supports monitoring and oversight of software experiments on computing clusters. The setup is particularly valuable for administrators, as it provides an 
overview of cluster-wide energy consumption and resource utilization. For researchers, the tool is most beneficial when studying (large-scale) distributed applications, where aggregated metrics and dashboard views capture meaningful system behavior. In contrast, for experiments focused on single processes, finer-grained instrumentation is 
required, as the stack’s aggregation and visualization features introduce unnecessary overhead for such narrow-scope tasks.

For data collection, most metrics are collected automatically by exporters running on the compute nodes, including CPU, memory, and power consumption. This ensures reproducibility and minimizes manual intervention during experiments.
For visualization, Grafana dashboards provide ready-to-use templates giving  
researchers an immediate starting point for experiment monitoring. 
Real-time views are available during execution, while aggregated dashboards summarize node-level utilization and efficiency for administrators. In addition, Grafana’s interface allows researchers to export selected data segments
, enabling them to perform custom analyzes and integrate the results into their 
workflows.

\begin{figure}
    \centering
    \includegraphics[width=0.85\linewidth]{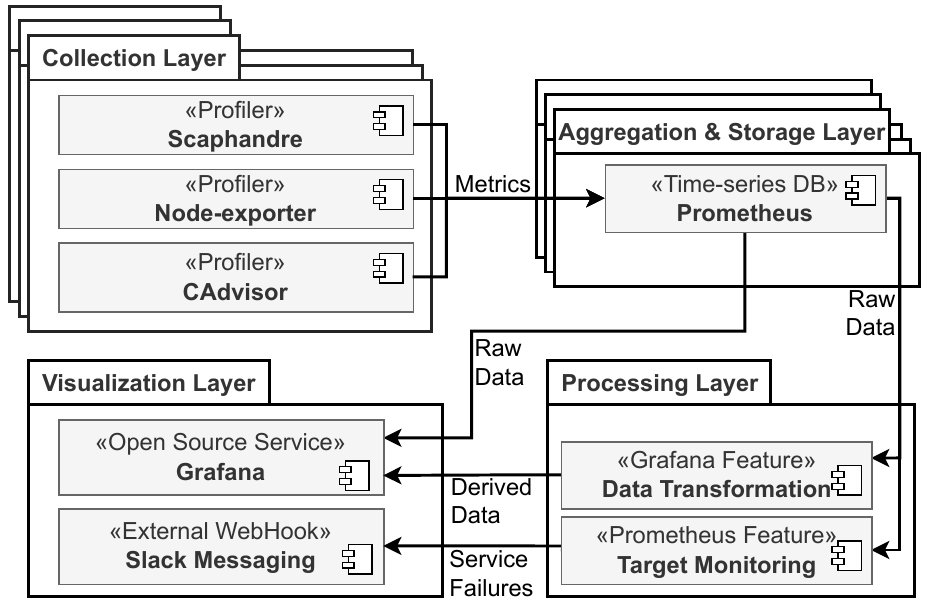}
    \caption{The GreenLab distributed monitoring stack}
    \label{fig:greenlabstack}
\end{figure}

The specifics of the current architecture are illustrated in Figure \ref{fig:greenlabstack}. The specificity of this use case is that the collection and the aggregation and storage layers are distributed. 
The code for this use case can also be found on GitHub~\cite{PeekStackgithub}.

\section{Conclusion and Future Work}
\label{sec:futurework}
\label{sec:conclusion}
The proposed architectural blueprint and associated code base toolkit provides support and examples for researchers and practitioners who would like to add sustainability-oriented observability to their (software) systems but are not proficient in how to build such observability stacks. In addition, the two use cases described propose comprehensive open-source examples of how the blueprint can be used in practice. 

While our current work introduces a reusable observability stack and demonstrates its applicability in two use cases, several directions remain open for further work.
Systematic \textit{validation studies} are required to assess the effectiveness of our blueprint across contexts, evaluating usability and adaptability for researchers, administrators, non-IT users, as well as its support for decision-making driven by sustainability metrics.
Additionally, the modular nature of the stack highlights opportunities for creating a curated ecosystem of \textit{domain-specific collector and visualization modules}, tailored to domains like agriculture, education, or HPC, to reduce deployment time and foster reproducibility. At the same time, challenges remain in ensuring accurate and efficient sustainability-oriented observability: metric isolation, profiling overhead in distributed environments, and clear documentation of profiler limitations should be addressed to improve reliability and adoption.  

\section*{Acknowledgment}
The authors would like to thank Kostas Vilkelis for his contribution to the initial design and implementation of the Feed4Food use case. Klervie Toczé and Patricia Lago were supported by the Feed4Food project, Grant Agreement n° 101069506, which has been funded by: Research and Innovation Foundation (Cyprus), Dutch Research Council (Netherlands)
General Secretariat for Research and Innovation (Greece), UE Fiscdi (Romania) under the Driving Urban Transitions Partnership, co-funded by the European Union.

\bibliographystyle{IEEEtran}
\bibliography{sample}

\end{document}